\documentclass[epj,nopacs,final]{svjour}
\usepackage{graphicx}
\usepackage{amsmath}
\usepackage{amssymb}

\begin{document}

\title{Thermodynamic anomaly of the free damped quantum particle:
the bath perspective}
\author{Gert-Ludwig Ingold
\thanks{\email{gert.ingold@physik.uni-augsburg.de}}
}
\institute{Institut f\"ur Physik, Universit\"at Augsburg, Universit\"atsstra{\ss}e 1,
86135 Augsburg, Germany}
\date{\today}

\abstract{%
A possible definition of the specific heat of open quantum systems is based on
the reduced partition function of the system. For a free damped quantum
particle, it has been found that under certain conditions, this specific
heat can become negative at low temperatures. In contrast to the conventional
approaches focusing on the system degree of freedom, here we concentrate on
the changes induced in the environment when the system is coupled to it. Our
analysis is carried out for an Ohmic environment consisting of harmonic
oscillators and allows to identify the mechanism by which the specific heat
becomes negative. Furthermore, the formal condition for the occurrence of
a negative specific heat is given a physical interpretation in terms of the
total mass of bath oscillators and the system mass.}

\maketitle

\section{Introduction}

The standard definition of the thermodynamic equilibrium quantities in a canonical
ensemble relies on the assumption that the coupling between the system of
interest and its thermal environment is arbitrarily weak. In the real world,
this approximation never strictly holds and important deviations from the usual
picture can be expected in particular for systems on the nanoscale. Beyond
the study of effects of finite coupling to the environment on the thermodynamic
properties of system, the more fundamental question arises of how to properly
define quantities like an internal energy, a specific heat or an entropy in 
such a situation \cite{hangg06}. The last five years or so have seen considerable
activities addressing this issue and related ones 
\cite{ford07,hangg08,hoerh08,wang08,ingol09,campi09a,campi09,gelin09,%
bandy09,kumar09,campi10,datta10,bandy10,bandy10a,kim10,campi11,willi11,haseg11b}.

In the quantum regime, the thermodynamic equilibrium properties not only depend
on the strength of the coupling to the environment. Even more importantly, it
turns out that the very definition of these quantities is no longer unique
\cite{hangg06}. As an example, we consider here the specific heat which for a
damped quantum system can be defined in at least two different ways. 

The first approach defines the internal energy of the system as expectation
value of the system Hamiltonian taken with respect to the thermal ensemble of
the coupled system. By the latter we mean here and in the following the coupled
complex of system (S) and heat bath (B). Quantities related to the coupled
system will be indicated by a subscript ``S+B''. Based on the definition of the
internal energy just introduced, one obtains the associated specific heat by
taking the derivative with respect to temperature. Obviously, in the absence
of a coupling between system and bath, the specific heat defined in this way
reduces to the standard specific heat of the system.

An alternative approach is based on the reduced partition function of the
system defined as \cite{dittr98,weiss99,ingol02}
\begin{equation}
\label{eq:reducedPartitionFunction}
\mathcal{Z} = \frac{\mathcal{Z}_\mathrm{S+B}}{\mathcal{Z}_\mathrm{B}}\,.
\end{equation}
In the absence of a coupling between the system and its environment, $\mathcal{Z}$
equals the partition function of the system. Even though we thus can
consider $\mathcal{Z}$ to be a quantity associated with the
system, we omit a subscript ``S'' in order to emphasize the fact that 
$\mathcal{Z}$ also takes into account the coupling between system and bath.
Here, we will employ the ratio (\ref{eq:reducedPartitionFunction}) only in
the context of equilibrium thermodynamics, but recently it has also been used in the
discussion of fluctuation theorems for open quantum systems in nonequilibrium
situations \cite{campi09a,campi09,campi11}.

The partition function (\ref{eq:reducedPartitionFunction}) can be used to
define thermodynamic equilibrium quantities by means of the standard relations
valid in the absence of a coupling between the system and its environment. We
thus obtain a second expression for the specific heat reading
\begin{equation}
\label{eq:csfromzred}
C = k_\mathrm{B}\beta^2
    \frac{\partial^2\ln(\mathcal{Z})}{\partial\beta^2}\,,
\end{equation}
where $\beta=1/k_\mathrm{B}T$. In the absence of a coupling between system and
bath, this definition of a specific heat also reduces to the usual definition
and thus agrees with our first definition in the limit of vanishing coupling.
Again we omit a subscript ``S'' to underline the dependence of the specific
heat on the coupling, but we do so also in view of the interpretation
(\ref{eq:diffcs}) given below which emphasizes the role of the heat bath.

It should be noted that, in general, the two specific heats just defined differ
for finite coupling to the environment already in the leading high-temperature
corrections to the classical specific heat. For a bilinear coupling between
system and bath, the classical specific heats still agree
\cite{hangg06,hangg08} while this need no longer be the case for anharmonic
couplings \cite{gelin09}.

The difference between the two approaches is most spectacular
in situations where the specific heat (\ref{eq:csfromzred}) becomes negative in
contrast to the positive specific heat based on the expectation value of the
system Hamiltonian \cite{hangg06,hangg08}. In the following, we exclusively consider
the specific heat obtained from the reduced partition function so that
the symbol $C$ will always refer to (\ref{eq:csfromzred}). 

Although the reduced partition function (\ref{eq:reducedPartitionFunction})
appears to be a rather formal starting point, the quantities derived from it
have a clear physical meaning. Due to the fact that thermodynamic equilibrium
properties depend on the logarithm of the partition function, these quantities
in the presence of a finite coupling to a heat bath possess a natural
interpretation in terms of the difference between the quantity evaluated for
the coupled system and the quantity evaluated for the bath alone
\cite{ingol09,ford85}. For the specific heat (\ref{eq:csfromzred}), we thus 
have
\begin{equation}
\label{eq:diffcs}
C = C_\mathrm{S+B}-C_\mathrm{B}\,.
\end{equation}
The two quantities involved in this difference, $C_\mathrm{S+B}$ and
$C_\mathrm{B}$, constitute measurable quantities. Even though one may invoke a
superbath imposing the temperature on the bath and on the coupled system, the
coupling to the superbath may be assumed to be negligible. Therefore, we are
certainly allowed to employ the standard thermodynamic definition for the
quantities on the right-hand side of (\ref{eq:diffcs}). $C$ thus is not a
specific heat in the proper sense but it is the natural replacement for a
specific heat in a situation where the system of interest is coupled with 
non-negligible strength to an environment. The approach discussed here was
recently employed in the analysis of thermal data obtained for a metal
phosphate compound \cite{swain11}.  For the sake of simplicity, we will
occasionally refer to $C$ as specific heat of the system. In doing so, we
should however keep in mind the preceding discussion.

With the above interpretation of the specific heat (\ref{eq:csfromzred}),
negative values should not give rise to concerns with respect to thermodynamic
instabilities. In fact, there is no reason why the difference of two positive
quantities should be positive. The situation considered here should therefore
be distinguished from the negative specific heat appearing e.g. in self-gravitating
systems within a microcanonical description \cite{lynde68,thirr70,thirr03}.

Apart from the free damped quantum particle which we focus on in this paper,
negative entropies and/or specific heats have been discussed e.g. in the
context of the Casimir effect
\cite{klimc06,milto08,hoye07,ingol09a,intra09,canag10a,borda10a,rodri11}, Kondo
systems \cite{flore04,zitko08}, XY spin chains \cite{campi10}, two-level
fluctuators \cite{campi09} and energy transport in proteins \cite{sulai10}.
While for a single harmonic oscillator coupled to an Ohmic bath
\cite{hangg06,hangg08} or to a single bath oscillator \cite{ingol09} the
specific heat was found to be positive, this is not necessarily the case if a
system consisting of several harmonic oscillators is coupled to a finite
environment \cite{haseg11b}.

In the treatment of dissipative systems, the system degree of freedom usually
is in the focus of interest. On the other hand, the difference
(\ref{eq:diffcs}) of specific heats, even though conceptually associated with
the system degree of freedom, provides the motivation to take a different point
of view, namely the one of the environment. The question answered by the
specific heat (\ref{eq:diffcs}) really is: How does the specific heat of the
environment change when a system degree of freedom is coupled to it?

In the following we shall address this question by studying the change in the
spectral density of the environment when a system degree of freedom is coupled
to it.  In Section~\ref{sec:freeParticleHarmonicOscillators} we briefly
introduce the model for the damped free quantum particle which will allow us to
obtain the change in the spectral density of the environment in
Section~\ref{sec:changeSpectralDensity}. From these results, we derive in
Section~\ref{sec:thermodynamicProperties} the thermodynamic properties of the
system in the sense explained above. We will show that these results can indeed
be expressed in terms of properties of the system degree of freedom. In
Section~\ref{sec:missingMass} we will give a physical interpretation of the
condition under which the specific heat (\ref{eq:diffcs}) of the free damped
quantum particle becomes negative. Finally, we present our conclusions in
Section~\ref{sec:conclusions}.

\section{Free particle coupled to harmonic oscillators}
\label{sec:freeParticleHarmonicOscillators}

As our model for the study of the appearance of a negative specific heat we consider
a particle of mass $M$ which is bilinearly coupled to a set of harmonic oscillators
constituting the environment. The corresponding Hamiltonian is given by
\begin{equation}
\label{eq:hamiltonian}
H = \frac{P^2}{2M} + \sum_{n=1}^\infty\left[\frac{p_n^2}{2m_n}+\frac{m_n\omega_n^2}{2}
(Q-x_n)^2\right]\,.
\end{equation}
The sum describes the environmental oscillators of mass $m_n$ and frequency $\omega_n$,
the bilinear coupling and a potential renormalization. The latter ensures that the
effective equation of motion of the system position $Q$ corresponds indeed to
the Langevin equation of a free damped particle taking the form
\begin{equation}
\label{eq:LangevinEquation}
M\ddot Q+M\int^t\mathrm{d}s\gamma(t-s)\dot Q(s) = \xi(t)\,.
\end{equation}
Below, we will relate the damping kernel $\gamma(t)$ to the properties of the
environment. The properties of the noise term $\xi(t)$ are irrelevant for our
discussion so that we refer the reader to the literature for details
\cite{dittr98,weiss99,ingol02,hangg05}.  The coupling constant of the bilinear
term in $x_n$ and $Q$ in (\ref{eq:hamiltonian}) has been expressed in terms of
the masses and frequencies of the environmental oscillators which is possible
without loss of generality \cite{hakim85,grabe88}. The Hamiltonian is manifestly
invariant under spatial translations of all degrees of freedom, confirming once
more that we are treating a free damped particle.

In our discussion, two spectral densities play a key role and it is important
to carefully distinguish them. We refer to the first quantity as spectral
density of eigenmodes $\rho$. This quantity is defined in terms of the
eigenfrequencies of the quadratic Hamiltonian (\ref{eq:hamiltonian}). For the 
bath alone, we have the spectral density of eigenmodes
\begin{equation}
\label{eq:spectralDensityOfEigenmodes0}
\rho_\mathrm{B}(\omega) = \sum_{n=1}^\infty\delta(\omega-\omega_n)\,.
\end{equation}
Correspondingly, we define a spectral density of eigenmodes
$\rho_\mathrm{S+B}(\omega)$ in which the frequencies $\omega_n$ appearing in
(\ref{eq:spectralDensityOfEigenmodes0}) are replaced by the eigenfrequencies of
the coupled system described by (\ref{eq:hamiltonian}). The difference 
$\rho_\mathrm{S+B}-\rho_{B}$ will form the basis for the derivation of various
thermodynamic quantities in Section~\ref{sec:thermodynamicProperties}.

In addition, in order to characterize the properties of the heat bath and its
coupling to the system, a quantity commonly referred to as spectral density
of bath oscillators \cite{calde83}
\begin{equation}
\label{eq:spectralDensityOfBathOscillators}
J(\omega) = \frac{\pi}{2}\sum_{n=1}^\infty m_n\omega_n^3\delta(\omega-\omega_n)
\end{equation}
is introduced. In contrast to the spectral density of eigenmodes
(\ref{eq:spectralDensityOfEigenmodes0}) in the absence of coupling, the spectral
density of bath oscillators (\ref{eq:spectralDensityOfBathOscillators}) not
only depends on the frequencies of the bath oscillators but also on their
masses. For our choice of coupling constants (cf. discussion below
(\ref{eq:LangevinEquation})) this is tantamount to saying that
(\ref{eq:spectralDensityOfBathOscillators}) depends also on the coupling
constants between system degree of freedom and bath oscillators.  $J(\omega)$
contains all information about the heat bath required to describe the
properties of the damped quantum system.

Although it is not obvious, it will turn out that even within our bath-centered
approach all quantities of interest will eventually be expressible in terms of
the spectral density of bath oscillators
(\ref{eq:spectralDensityOfBathOscillators}). This already indicates that even
though we take the point of view of the environment, we will learn something
about the properties of the system.

In our results, it will be natural to express the dependence on $J(\omega)$ in
terms of the Laplace transform of the damping kernel $\gamma(t)$ appearing in
(\ref{eq:LangevinEquation}). The two quantities are related by means of
\cite{dittr98,weiss99,ingol02}
\begin{equation}
\label{eq:gammaHat}
\hat\gamma(z) = \frac{2}{\pi M}\int_0^\infty\mathrm{d}\omega\frac{J(\omega)}{\omega}
\frac{z}{\omega^2+z^2}\,.
\end{equation}

For the purpose of our discussion, we will restrict ourselves to so-called Ohmic
damping which implies that the spectral density of bath oscillators is continuous
and linear in the frequency at least for small frequencies. At high frequencies,
the spectral density of bath oscillators will typically be suppressed with respect
to this linear behavior. To be specific, in cases where such a cutoff 
is of relevance, we will employ the so-called Drude model with
\begin{equation}
\label{eq:densityDrude}
J_\mathrm{D}(\omega) = M\gamma\omega
                     \frac{\omega_\mathrm{D}^2}{\omega^2+\omega_\mathrm{D}^2}\,.
\end{equation}
Here, the low-frequency behavior is characterized by the damping constant $\gamma$
and $\omega_\mathrm{D}$ is the cutoff frequency. The corresponding Laplace transform 
of the damping kernel reads
\begin{equation}
\label{eq:gammaHatDrude}
\hat\gamma_\mathrm{D}(z) = \frac{\gamma\omega_\mathrm{D}}{z+\omega_\mathrm{D}}\,.
\end{equation}
Because the damping constant $\gamma$ combines with the inverse temperature 
$\beta$ to form a dimensionless quantity $\hbar\beta\gamma$ and thus
merely sets the temperature scale, the cutoff frequency $\omega_\mathrm{D}$ will be
an important parameter determining the properties of the damped free particle.

\section{Change in the spectral density of bath oscillators}
\label{sec:changeSpectralDensity}

In our environment-centered approach, we start out with a set of uncoupled environmental
oscillators with frequencies $\omega_n$. In addition, we have one system degree of freedom
corresponding to an undamped free particle. We now couple the environmental oscillators
to the free particle as prescribed by the Hamiltonian (\ref{eq:hamiltonian}). As discussed
in the previous section, the system and environment together are translationally invariant.

In the spirit of Ullersma's analysis of the damped harmonic oscillator
\cite{uller66} we determine the eigenmode spectrum of the free particle coupled
to its environment.  Diagonalizing the Hamiltonian, we recover a zero-mode
which replaces the zero-mode corresponding to the uncoupled free particle. We
thus concentrate on the non-zero frequencies which due to the coupling are
shifted with respect to the original environmental oscillator frequencies.  The
new frequencies are obtained as solutions $\Omega$ of the equation
\begin{equation}
\label{eq:evCondition}
\sum_{n=1}^\infty \frac{m_n\omega_n^2}{\Omega^2-\omega_n^2} = M\,.
\end{equation}

For the following intermediate steps it is convenient to explicitly consider an
equidistant set of discrete frequencies $\omega_n=n\Delta$ with $n=1, 2, \dots$
for the environmental oscillators. Later, we will take the limit of vanishing frequency
spacing $\Delta$ to recover a continuous spectral density of environmental oscillators.
In view of (\ref{eq:spectralDensityOfBathOscillators}), we have the relation
\begin{equation}
\int_0^\infty\mathrm{d}\omega\,J(\omega)f(\omega) = \frac{\pi}{2}\sum_{n=1}^\infty
m_n\omega_n^3f(n\Delta)
\end{equation}
which holds for any function $f$ for which the integral and the sum exist. If, on the 
other hand, we replace the integral on the left-hand side by a Riemann sum with step
width $\Delta$, we can express the masses $m_n$ of the environmental oscillators in terms
of the continuous spectral density of bath oscillators $J(\omega)$ according to
\begin{equation}
m_n = \frac{2}{\pi}\frac{J(n\Delta)}{(n\Delta)^3}\Delta\,.
\end{equation}
The eigenfrequency condition (\ref{eq:evCondition}) thus becomes
\begin{equation}
\sum_{n=1}^\infty\frac{J(n\Delta)}{n[\Omega^2-(n\Delta)^2]} = \frac{\pi}{2}M\,.
\end{equation}
In order to treat frequencies $\Omega$ close to an unperturbed environmental frequency
$n\Delta$ correctly, we rewrite this condition as
\begin{equation}
\begin{aligned}
&\Delta\frac{J(\Omega)}{\Omega}\sum_{n=1}^\infty\frac{1}{\Omega^2-(n\Delta)^2}\\
&\quad+\Delta\sum_{n=1}^\infty\frac{1}{\Omega^2-(n\Delta)^2}\left(\frac{J(n\Delta)}{n\Delta}
-\frac{J(\Omega)}{\Omega}\right)=\frac{\pi}{2}M
\end{aligned}
\end{equation}
Performing the first sum exactly and replacing the second sum by an integral, we finally
obtain
\begin{equation}
\label{eq:newEigenvaluesCondition}
\cot\left(\frac{\pi\Omega}{\Delta}\right)-\frac{\Delta}{\pi\Omega} = g(\Omega)
\end{equation}
with
\begin{equation}
\label{eq:functiong}
g(\Omega) = \frac{M\Omega}{J(\Omega)}\big(\Omega+\mathrm{Im}\hat\gamma(\mathrm{i}\Omega)\big)\,.
\end{equation}

The coupling of the environment to the system degree of freedom modifies the
original spacing $\Delta$ between adjacent environmental eigenfrequencies
yielding a new spacing $\Delta+\epsilon$.  The correction $\epsilon$ depends on
the spectral bath density via the function $g$ defined in (\ref{eq:functiong})
and can be shown to be of order $\Delta^2$. Exploiting the latter fact and
making use of the addition theorem of the cotangent, one determines $\epsilon$
from (\ref{eq:newEigenvaluesCondition}). For the change in the spectral density
of eigenmodes, one thus finds as a central result
\begin{equation}
\label{eq:rhoMinusRho0}
\rho_\mathrm{S+B}(\Omega)-\rho_\mathrm{B}(\Omega) = 
\frac{1}{\Delta+\epsilon(\Omega)}-\frac{1}{\Delta}
= \frac{1}{\pi}\frac{g'(\Omega)}{1+g(\Omega)^2}\,.
\end{equation}
Here, the prime denotes a derivative with respect to the argument.

Further insight into the change of the spectral density of eigenmodes can be
obtained by specifying the properties of the bath. We choose a Drude model 
with the spectral density of environmental oscillators given by 
(\ref{eq:densityDrude}) and a Laplace transform of the damping kernel as specified
in (\ref{eq:gammaHatDrude}). From (\ref{eq:rhoMinusRho0}) one obtains
\begin{equation}
\label{eq:lorentziansReal}
\rho_\mathrm{S+B}-\rho_\mathrm{B} = \frac{1}{\pi}\left[\frac{\omega_1}{\Omega^2+\omega_1^2}
+\frac{\omega_2}{\Omega^2+\omega_2^2}
-\frac{\omega_\mathrm{D}}{\Omega^2+\omega_\mathrm{D}^2}\right]
\end{equation}
where
\begin{equation}
\label{eq:omega12}
\omega_{1,2} = \omega'\pm\mathrm{i}\omega''
= \frac{\omega_\mathrm{D}}{2}\left(1\pm\sqrt{1-\frac{4\gamma}{\omega_\mathrm{D}}}\right)
\end{equation}
and $\omega_\mathrm{D}$ is the Drude frequency providing the high-fre\-quency cutoff.
These three frequencies are the eigenfrequencies associated with the deterministic
version of the Langevin equation (\ref{eq:LangevinEquation}) for a Drude damping
kernel \cite{hangg08}. 

\begin{figure}
\begin{center}
\includegraphics[width=0.9\columnwidth]{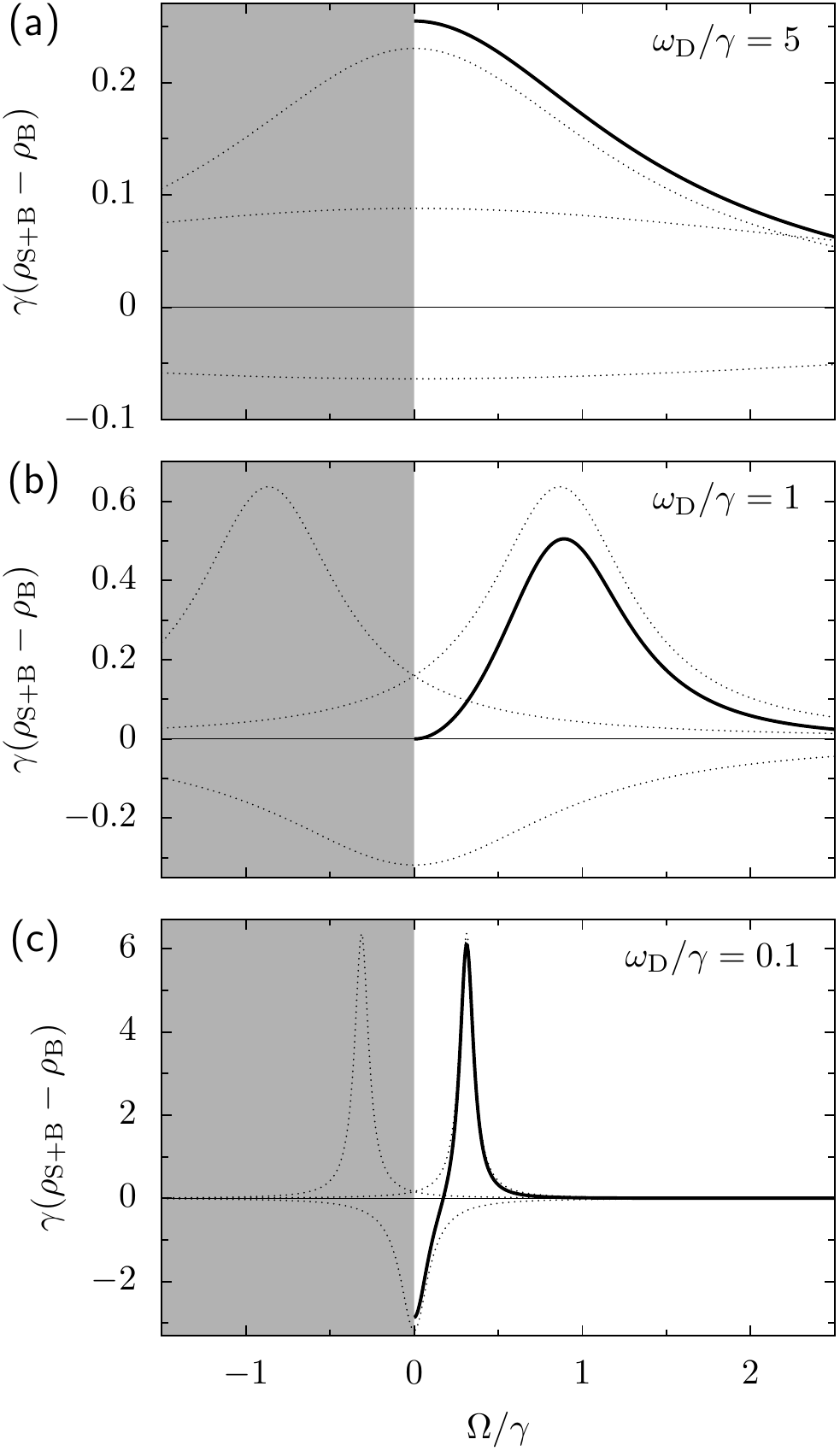}
\end{center}
\caption{The change in the spectral density of eigenmodes
$\rho_\mathrm{S+B}-\rho_\mathrm{B}$ induced by the coupling to the system degree 
of freedom is depicted as thick solid line for Drude damping with (a) 
$\omega_\mathrm{D}/\gamma=5$, (b) $\omega_\mathrm{D}/\gamma=1$,
and (c) $\omega_\mathrm{D}/\gamma=0.1$. The contributions of the three Lorentzians 
according to (\ref{eq:lorentziansReal}) are represented as dotted lines. Only
the white regions of positive frequencies are of relevance for the spectral
density of eigenmodes.}
\label{fig:rho_lorentz}
\end{figure}

The change in the spectral density of eigenmodes is shown as thick solid line in
Fig.~\ref{fig:rho_lorentz} for various values of $\omega_\mathrm{D}/\gamma$. 
The dotted lines represent the Lorentzian contributions according to 
(\ref{eq:lorentziansReal}). For the sake of clarity, they are depicted also in 
the grey regions of negative frequencies which are irrelevant for the spectral
density of eigenmodes.  As long as the cutoff frequency is sufficiently high, 
$\omega_\mathrm{D}>4\gamma$, (\ref{eq:lorentziansReal}) describes the sum of 
three Lorentzians centered at zero frequency. An example for this situation is 
given in Fig.~\ref{fig:rho_lorentz}a. Coupling the system degree of freedom to the
environment then leads to an increase of the spectral density at zero frequency
by $(\omega_\mathrm{D}-\gamma)/\pi\gamma\omega_\mathrm{D}$.  For smaller cutoff
frequencies, if $\omega_\mathrm{D}<4\gamma$ like in the cases depicted in 
Figs.~\ref{fig:rho_lorentz}b and c, the two frequencies (\ref{eq:omega12}) become
complex and the first two Lorentzians in (\ref{eq:lorentziansReal}) are now
centered at nonzero frequencies.  As a consequence, the suppressing effect of
the third Lorentzian on the spectral density of eigenmodes becomes relevant.
For $\omega_\mathrm{D}<\gamma$ one finds in fact a suppression of the spectral
density at zero frequency as is shown in Fig.~\ref{fig:rho_lorentz}c for
$\omega_\mathrm{D}/\gamma=0.1$. It is this suppression which leads to the
negative specific heat as we will discuss in the following section. 

Before doing so, we would like to emphasize that although the change in the
spectral density of eigenmodes obtained in this section is mainly related to
properties of the environment, the results are determined by properties of the
damped system. In the general result (\ref{eq:rhoMinusRho0}) the environment
appears only through the spectral density of bath oscillators
(\ref{eq:spectralDensityOfBathOscillators}) which is sufficient to provide a
reduced description of the damped system. In the specific case of Drude
damping, the three eigenfrequencies of the damped system turn out to be
sufficient for the complete description of the change in the spectral density
of eigenfrequencies.

Despite being closely related to the properties of the damped system, the
change of the spectral density of eigenmodes discussed here should not be
confused with the density of states which can be obtained from the reduced
partition function (\ref{eq:reducedPartitionFunction}) by its inverse Laplace
transform \cite{hanke95}. Because the density of states defined in such a way
is linearly related to the reduced partition function, it cannot be interpreted
as a difference of a property of the coupled system on the one hand and the
heat bath on the other hand \cite{hangg08}. 

\section{Thermodynamic properties of the system}
\label{sec:thermodynamicProperties}

The change of the specific heat (\ref{eq:csfromzred}) due to the coupling of
the system degree of freedom to the environment can be obtained from the change
(\ref{eq:rhoMinusRho0}) of the spectral density of eigenfrequencies as
\begin{equation}
\label{eq:csGeneral}
C = \int_0^\infty\mathrm{d}\omega
\big(\rho_\mathrm{S+B}(\omega)-\rho_\mathrm{B}(\omega)\big)C_\mathrm{ho}(\omega)
\end{equation}
where
\begin{equation}
\label{€q:cho}
C_\mathrm{ho}(\omega) = k_\mathrm{B}
\left(\frac{\hbar\beta\omega}{2\sinh(\hbar\beta\omega/2)}\right)^2\,.
\end{equation}
is the specific heat of a harmonic oscillator with frequency $\omega$.
Eq.~\ref{eq:csGeneral} implies a negative specific heat $C$ at low temperatures
if the spectral density of eigenmodes is suppressed for small frequencies due
to the coupling of the system degree of freedom to the bath.

Assuming Ohmic damping, i.\,e. $J(\omega)\sim\omega$ at low frequencies,
we can obtain the specific heat at low temperatures by inserting the change
in the spectral density of eigenmodes (\ref{eq:rhoMinusRho0}) into 
(\ref{eq:csGeneral}). Together with (\ref{eq:functiong}) one finds
\begin{equation}
\label{eq:csLowTemperature}
\frac{C}{k_\mathrm{B}} = \frac{\pi}{3}\frac{1+\hat\gamma^\prime(0)}
{\hat\gamma(0)}\frac{k_\mathrm{B}T}{\hbar} + \mathcal{O}(T^3)
\end{equation}
in agreement with the findings of Ref.~\cite{hangg08}. For 
\begin{equation}
\label{eq:negativityCondition}
\hat\gamma^\prime(0) <-1\,, 
\end{equation}
we thus obtain a negative specific heat at low temperatures. A physical
interpretation of this result will be given in Section~\ref{sec:missingMass} below.

More explicit results can be derived for the case of Drude damping. There, the 
integral in (\ref{eq:csGeneral}) can be evaluated analytically and one finds
\begin{equation}
\begin{aligned}
\label{eq:csDrude}
\frac{C}{k_\mathrm{B}} &= 
\left(\frac{\hbar\beta\omega_1}{2\pi}\right)^2\psi^\prime
  \left(\frac{\hbar\beta\omega_1}{2\pi}\right)
+\left(\frac{\hbar\beta\omega_2}{2\pi}\right)^2\psi^\prime
  \left(\frac{\hbar\beta\omega_2}{2\pi}\right)\\
&\qquad-\left(\frac{\hbar\beta\omega_\mathrm{D}}{2\pi}\right)^2\psi^\prime
  \left(\frac{\hbar\beta\omega_\mathrm{D}}{2\pi}\right)
-\frac{1}{2}\,,
\end{aligned}
\end{equation}
where $\psi^\prime(z)$ denotes the digamma function. This result is in
agreement with the expression obtained by proceeding according to the point of
view of the system \cite{hangg08} and starting with the reduced partition function
(\ref{eq:reducedPartitionFunction}). The two approaches therefore are equivalent,
but the approach presented here gives additional insight through the change
in the spectral density of eigenvalues (\ref{eq:rhoMinusRho0}). 

Instead of obtaining the internal energy and the specific heat from the reduced
partition function, we can use our result (\ref{eq:csDrude}) for the specific
heat to obtain the other two quantities. In the limit of vanishing coupling to
the heat bath, the internal energy $U$ is related to the specific heat $C$ by means of 
\begin{equation}
C=-\beta^2\frac{\partial U}{\partial\beta}\,.
\end{equation}
In view of (\ref{eq:diffcs}) we thus obtain from (\ref{eq:csDrude}) by means of an
integration the difference of internal energies induced by the system-bath coupling
\begin{equation}
\begin{aligned}
U &= U_\mathrm{S+B}-U_\mathrm{B}\\
&= -\frac{\hbar\omega_1}{2\pi}\psi\left(\frac{\hbar\beta\omega_1}{2\pi}\right)
-\frac{\hbar\omega_2}{2\pi}\psi\left(\frac{\hbar\beta\omega_2}{2\pi}\right)\\
&\qquad+\frac{\hbar\omega_\mathrm{D}}{2\pi}\psi\left(\frac{\hbar\beta\omega_\mathrm{D}}{2\pi}\right)
-\frac{1}{2\beta}\,.
\end{aligned}
\end{equation}
This result is only determined up to a constant of integration which, by 
comparison with the known result \cite{hangg08}, turns out to vanish. In 
particular, at zero temperature, we thus find
\begin{equation}
U_0 = \frac{\hbar\omega_1}{2\pi}\ln\left(\frac{\omega_\mathrm{D}}{\omega_1}\right)
+\frac{\hbar\omega_2}{2\pi}\ln\left(\frac{\omega_\mathrm{D}}{\omega_2}\right)\,.
\end{equation}

Furthermore, by means of the relation between the internal energy $U$ and the partition 
function $\mathcal{Z}$
\begin{equation}
U = -\frac{\mathrm{d}}{\mathrm{d}\beta}\ln(\mathcal{Z})
\end{equation}
one reproduces the correct temperature dependence of the ratio of the partition functions
of system and bath on the one hand and bath alone on the other hand \cite{hangg08}
\begin{equation}
\mathcal{Z}\sim\frac{1}{\beta^{1/2}}
\frac{\Gamma\left(1+\dfrac{\hbar\beta\omega_1}{2\pi}\right)
      \Gamma\left(1+\dfrac{\hbar\beta\omega_2}{2\pi}\right)}
{\Gamma\left(1+\dfrac{\hbar\beta\omega_\mathrm{D}}{2\pi}\right)}\,.
\end{equation}
This result leaves a prefactor undetermined which for the thermodynamical
equilibrium quantities is irrelevant. The temperature dependence, however,
agrees with the result obtained by means of other techniques, e.g. the
path integral approach.

\section{The missing mass of bath oscillators}
\label{sec:missingMass}

In the previous section, we have found that a negative specific heat occurs
provided the condition (\ref{eq:negativityCondition}) is satisfied.
This rather formal condition can be given a physical meaning. In order to
avoid an infrared divergence in the limit of vanishing argument $z$, we first
express the Laplace transform of the damping kernel (\ref{eq:gammaHat}) as
\begin{equation}
\label{eq:gammaHatDiff}
\hat\gamma(z) = \hat\gamma(0)+\frac{2}{\pi}\int_0^\infty\mathrm{d}\omega
\left(\frac{J(\omega)}{M\omega}-\hat\gamma(0)\right)\frac{z}{\omega^2+z^2}\,.
\end{equation}
With (\ref{eq:negativityCondition}) and (\ref{eq:gammaHatDiff}) the condition 
for a negative specific heat then becomes
\begin{equation}
\label{eq:negativityCondition2}
\frac{2}{\pi}\int_0^\infty\mathrm{d}\omega\frac{M\hat\gamma(0)\omega-J(\omega)}
{\omega^3}>M\,.
\end{equation}

Observing that according to (\ref{eq:spectralDensityOfBathOscillators}) the total
mass of the bath oscillators can be obtained from the spectral density of bath
oscillators $J(\omega)$ as \cite{hakim85,grabe88}
\begin{equation}
\mathcal{M} = \sum_{n=1}^\infty m_n = \frac{2}{\pi}\int_0^\infty\mathrm{d}\omega
\frac{J(\omega)}{\omega^3}
\end{equation}
the condition (\ref{eq:negativityCondition2}) can be expressed as
\begin{equation}
\label{eq:massCondition}
\Delta\mathcal{M}_\mathrm{B}>M\,.
\end{equation}
Here, $\Delta\mathcal{M}_\mathrm{B}$ is defined by the left-hand side of
(\ref{eq:negativityCondition2}) and refers to the total mass of oscillators
which are missing in the actual bath with respect to a strictly Ohmic reference
bath where $J(\omega)=M\hat\gamma(0)\omega$ for all frequencies. We note that 
while due to the infrared divergence mentioned above the total mass is infinite 
for both baths, the difference $\Delta\mathcal{M}_\mathrm{B}$ is finite.

We thus arrive at the following interpretation. Coupling a system degree of
freedom of zero frequency to the bath oscillators tends to shift the
frequencies of the latter to larger values. However, for a strictly Ohmic
environment, i.e. in the absence of a finite cutoff in the spectral density
of bath oscillators, the ensemble of bath oscillators resists a reduction of
the spectral density of eigenmodes at low frequencies. A suppression only becomes
possible if one eliminates a number of bath oscillator with a total mass at
least as large as the mass associated with the system degree of freedom. The
bath then is no longer able to resist the ``pressure'' of the system degree of
freedom pushing the eigenmodes to higher frequencies.

\section{Conclusions}
\label{sec:conclusions}

In order to achieve a better understanding of the unusual thermodynamic
properties of a free Brownian quantum particle, we have employed a somewhat
uncommon approach to the analysis of dissipative quantum systems by taking the
perspective of the environment. This approach is, however, quite natural if
thermodynamic quantities are defined in terms of a reduced partition function
of the system because they actually refer to the change of these quantities when
the system degree of freedom is coupled to the heat bath.

We have analyzed the modification of the bath spectrum induced by the coupling
to the system degree of freedom. Interestingly, the spectral density of bath
oscillators is sufficient to describe the change in the bath spectrum. As a
consequence the latter depends only on properties of the damped system as has
been exemplified by means of a Drude-type damping. Starting from the change in
the bath spectrum, expressions for the specific heat, the internal energy and
the reduced partition function obtained previously from a system-based approach
have been reproduced.

The low-temperature behavior of the specific heat of the free damped particle
is determined by the shift of the low-frequency environmental oscillators
induced by the coupling to the system degree of freedom. If the ratio of cutoff
frequency and damping strength is sufficiently small, the system degree of
freedom succeeds in suppressing the spectral density of eigenmodes at low
frequencies by shifting the bath modes to higher frequencies. As a result, at
low temperatures the specific heat of the heat bath is lowered if the system
degree of freedom is attached. In contrast, for larger cutoff frequencies, the
high-frequency oscillators of the environment act against the tendency of
the system to shift the bath modes to higher frequencies. Then, the specific
heat remains positive for all frequencies.

The condition to be satisfied by the environment to allow for a negative
specific heat at low temperatures has been shown to have a physical 
interpretation. The anomaly in the specific heat appears if the mass missing
in the environment with respect to a strictly Ohmic reference bath exceeds the
mass associated with the system degree of freedom. Then, the bath oscillators
are not strong enough to resist the zero-frequency degree of freedom which,
when coupled to the environment, tends to increase their frequencies. Otherwise
the spectrum of bath modes is sufficiently stiff to prevent the spectral
density of eigenmodes from being reduced.

\begin{acknowledgement}
The author is grateful to Michele Campisi, Peter H\"anggi, Astrid Lambrecht, 
Serge Reynaud, Peter Talkner, and Juan Diego Urbina for useful discussions. 
This work has been supported by the DAAD through the PROCOPE program.
\end{acknowledgement}


\begin{thebibliography}{99}

\bibitem{hangg06}
P. H\"anggi and G.-L. Ingold, Acta Phys. Pol. B \textbf{37}, 1537 (2006).

\bibitem{ford07}
G. W. Ford and R. F. O'Connell, Phys. Rev. B \textbf{75}, 134301 (2007).

\bibitem{hangg08}
P. H\"anggi, G.-L. Ingold, and P. Talkner, New J. Phys. \textbf{10}, 115008 (2008).

\bibitem{hoerh08}
C. H{\"o}rhammer and H. B{\"u}ttner, J. Stat. Phys. {\bf 133}, 1161 (2008).

\bibitem{wang08}
C.-Y. Wang and J.-D. Bao, Chin. Phys. Lett. \textbf{25}, 429 (2008).

\bibitem{ingol09}
G.-L. Ingold, P. H\"anggi, and P. Talkner, Phys. Rev. E \textbf{79}, 061105 (2009).

\bibitem{campi09a}
M. Campisi, P. Talkner, and P. H\"anggi, Phys. Rev. Lett. \textbf{102}, 210401 (2009).

\bibitem{campi09}
M. Campisi, P. Talkner, and P. H\"anggi,
J. Phys. A: Math. Theor. \textbf{42}, 392002 (2009).

\bibitem{gelin09}
M. F. Gelin and M. Thoss, Phys. Rev. E \textbf{79}, 051121 (2009).

\bibitem{bandy09}
M. Bandyopadhyay, J. Stat. Mech.: Theory Exp. P05002 (2009).

\bibitem{kumar09}
J. Kumar, P. A. Sreeram, and S. Dattagupta, Phys. Rev. E \textbf{79}, 021130 (2009).

\bibitem{campi10}
M. Campisi, D. Zueco, and P. Talkner, Chem. Phys. \textbf{375}, 187 (2010).

\bibitem{datta10}
S. Dattagupta, J. Kumar, S. Sinha, and P. A. Sreeram
Phys. Rev. E \textbf{81}, 031136 (2010).

\bibitem{bandy10}
M. Bandyopadhyay and S. Dattagupta, Phys. Rev. E \textbf{81}, 042102 (2010).

\bibitem{bandy10a}
M. Bandyopadhyay, J. Stat. Phys. \textbf{140}, 603 (2010).

\bibitem{kim10}
I. Kim and G. Mahler, Phys. Rev. E \textbf{81}, 011101 (2010).

\bibitem{campi11}
M. Campisi, P. H{\"a}nggi, and P. Talkner,
Rev. Mod. Phys. \textbf{83}, 771 (2011).

\bibitem{willi11}
N. S. Williams, K. Le Hur, and A. N. Jordan,
J. Phys. A: Math. Theor. \textbf{44}, 385003 (2011).

\bibitem{haseg11b}
H. Hasegawa, to appear in J. Math. Phys., \texttt{arXiv:1109.0108}.

\bibitem{dittr98}
T. Dittrich, P. H{\"a}nggi, G.-L. Ingold, B. Kramer, G. Sch{\"o}n, and W. Zwerger,
\textit{Quantum transport and dissipation}, Chap.\ 4 (Wiley-VCH, Weinheim, 1998).

\bibitem{weiss99}
U. Weiss, \textit{Quantum dissipative systems} (World Scientific, Singapore, 1999).

\bibitem{ingol02}
G.-L. Ingold, Lect. Notes Phys. \textbf{611}, 1 (2002).

\bibitem{ford85}
G. W. Ford, J. T. Lewis, and R. F. O'Connell, Phys. Rev. Lett. \textbf{55},
2273 (1985).

\bibitem{swain11}
T. Swain, J. Therm. Anal. Calorim. \textbf{103}, 1111 (2011).

\bibitem{lynde68}
D. Lynden-Bell and R. Wood, Mon. Not. R. Astr. Soc. \textbf{138}, 495 (1968).

\bibitem{thirr70}
W. Thirring, Z. Phys. \textbf{235}, 339 (1970).

\bibitem{thirr03}
W. Thirring, H. Narnhofer, and H. A. Posch,
Phys. Rev. Lett. \textbf{91}, 130601 (2003).

\bibitem{klimc06}
G. L. Klimchitskaya and V. M. Mostepanenko, Contemp. Phys. \textbf{47}, 131 (2006).

\bibitem{milto08}
K. A. Milton, J. Phys.: Conf. Ser. \textbf{161}, 012001 (2009).

\bibitem{hoye07}
J. S. H{\o}ye, I. Brevik, S. A. Ellingsen, and J. B. Aarseth,
Phys. Rev. E \textbf{75}, 051127 (2007).

\bibitem{ingol09a}
G.-L. Ingold, A. Lambrecht, and S. Reynaud, Phys. Rev. E \textbf{80}, 041113 (2009).

\bibitem{intra09}
F. Intravaia and C. Henkel, Phys. Rev. Lett. \textbf{103}, 130405 (2009).

\bibitem{canag10a}
A. Canaguier-Durand, P. A. Maia Neto, A. Lambrecht, and S. Reynaud,
Phys. Rev. A \textbf{82}, 012511 (2010).

\bibitem{borda10a}
M. Bordag and I. G. Pirozhenko, Phys. Rev. D \textbf{82}, 125016 (2010).

\bibitem{rodri11}
P. Rodriguez-Lopez, Phys. Rev. B \textbf{84}, 075431 (2011).

\bibitem{flore04}
S. Florens and A. Rosch, Phys. Rev. Lett. \textbf{92}, 216601 (2004).

\bibitem{zitko08}
R. {\v Z}itko and T. Pruschke, Phys. Rev. B \textbf{79}, 012507 (2009).

\bibitem{sulai10}
A. Sulaiman, F. P. Zen, H. Alatas, and L. T. Handoko,
Phys. Rev. E \textbf{81}, 061907 (2010).

\bibitem{hangg05}
P. H\"anggi and G.-L. Ingold, Chaos \textbf{15}, 026105 (2005).

\bibitem{hakim85}
V. Hakim and V. Ambegaokar, Phys. Rev. A \textbf{32}, 423 (1985).

\bibitem{grabe88}
H. Grabert, P. Schramm, and G.-L. Ingold, Phys. Rep. \textbf{168}, 115 (1988).

\bibitem{calde83}
A. O. Caldeira and A. J. Leggett, Ann. Phys. (N.Y.) \textbf{149}, 374 (1983).

\bibitem{uller66}
P. Ullersma, Physica \textbf{32}, 27 (1966).

\bibitem{hanke95}
A. Hanke and W. Zwerger, Phys. Rev. E \textbf{52}, 6875 (1995).

\end{thebibliography}
\end{document}